\documentclass[twocolumn,prb,showpacs,preprintnumbers,a4paper]{revtex4}
\usepackage{graphicx}
\usepackage{bm}
\begin{document}

\title{Density-functional study of oxygen adsorption on Mo(112)}

\author{Adam Kiejna$^1$ and R. M.\ Nieminen$^2$}
\affiliation{$^1$Institute of Experimental Physics, University of Wroc{\l}aw,
  Plac M. Borna 9, PL-50-204 Wroc{\l}aw, Poland\\ 
$^2$Laboratory of Physics, Helsinki University of Technology, 
 P.O.Box 1100, FIN-02015 HUT, Finland}

\date{\today}

\begin{abstract}
Atomic oxygen adsorption on the Mo(112) surface has been investigated by 
means of first-principles total energy calculations. Among the variety
of possible adsorption sites it was found that the bridge sites between 
two Mo atoms of the topmost row are favored for O adsorption at low and 
medium coverages.
At about one monolayer coverage oxygen atoms prefer to adsorb in a
quasi-threefold hollow sites coordinated by two first-layer Mo atoms and
one second layer atom. The stability of a structural model for an oxygen-induced
$p(2\times 3)$ reconstruction of the missing-row type is examined.
\end{abstract}


\maketitle

\section{INTRODUCTION}

Molybdenum surfaces attract substantial current interest because of
their important applications and fundamental properties.
Molybdenum is one of the most interesting catalysts for its
various oxidation states and a wide range of chemical reactivity.
Oxidation of Mo strongly alters the dissociative adsorption
of several molecules. Thus, oxygen adsorption on the low-index planes of
{\it bcc} Mo has been widely studied using various experimental techniques.
The preferred adsorption sites and ordered structures have been identified
(see Ref.~\onlinecite{Schro02} and references therein).

The Mo(112) surface although open, is also very stable and is exploited
as a substrate for the growth of ordered silica layers \cite{Schro02a,Chen04} 
or epitaxial growth of MoO$_2$.
The furrowed structure of Mo(112) has been used to study anisotropy of
interactions in adsorbed submonolayers of alkali \cite{KieNie02} and
alkaline-earth metal \cite{KieNie04} atoms. Recent experimental findings,
reporting that a presence of small amount of oxygen adatoms promotes a
longer periodicity of low-coverage linear-chain-structures \cite{KoPf_2,Godz04}
formed in the Mo(112) furrows, and responsible for commensurate
phase transition in adsorbed Sr layers, \cite{Godz01} motivated this
theoretical examination of O adsorption on Mo(112).

According to experimental studies of low-coverage O adsorption
\cite{Fuku93,Aruga95,Sas02} at Mo(112), oxygen atoms are expected to adsorb
in trough sites between close-packed Mo rows and the sequence of most stable
atomic oxygen sites can be ordered as: quasi-threefold hollow, long bridge,
and on-top.

Oxygen adsorption on the more open metal surfaces is known for long time
to induce their reconstruction. One of the best known examples is the
$(2\times 1)$ added-row reconstruction of the fcc Cu(110) surface
\cite{Ertl67,Lie98} which has a similar trough-and-row structure to the
bcc (112).
Different possibilities for O and C induced reconstructions of Mo(112)
and their effect on the surface band structure were discussed, \cite{McA00} 
but it is only recently when Schroeder \textit{et al.} \cite{Schro02,Schro04} 
reported results of
a detailed experimental study of the transformation of the clean Mo(112) 
surface under oxygen exposure into an epitaxially grown MoO$_2$ film. 
They found that an oxygen-induced $p(2\times 3)$ reconstruction precedes 
the MoO$_2$  formation. This phase is very stable and passivates the  
Mo(112) against further oxidation in its initial stages.

The principal goal of this investigation is the determination of the oxygen 
modification of a Mo(112) surface and identification of preferred adsorption 
sites during low-coverage atomic oxygen adsorption, varying from a small 
fraction of a monolayer (ML) up to one monolayer. The stability of different
sites and  configurations of O adatoms is explored including the observed 
$p(1\times 2)$ pairing-row overlayer structure which can be viewed as the
one-dimensional MoO$_2$ oxide structure. 
To the best of our knowledge such calculations have not been performed so 
far. We also investigate the stability of the structural model of the 
experimentally discovered oxygen induced $(2\times 3)$ missing-row
reconstruction of Mo(112). \cite{Schro02,Schro04}

The remaining part of the paper is organized as follows. Section II 
outlines the methodology of calculations. It is followed by a presentation 
and discussion of our results in Sec.~III. The paper is summarized in a 
concluding section.

\section{Computational Methods}

The calculations are based on density-functional theory \cite{HK64,KS65}
in the generalized gradient approximation \cite{Per92} for the 
exchange-correlation energy functional. The electron-ion interaction is
represented by the projector-augmented-wave potentials \cite{Blo94} as 
implemented \cite{paw} in the Vienna \textit{ab initio} simulation package
({\sc vasp}). \cite{vasp}
The ground state is determined by solving the Kohn-Sham equations applying 
a plane waves basis set with energy cutoff of 300 eV.
A clean Mo(112) surface is modeled by periodic slabs consisting of seven
molybdenum layers separated by empty space of sufficient thickness (eight 
equivalent Mo layers). Oxygen atoms are adsorbed in various sites on one 
side of the slab. The positions of Mo atoms in the four topmost layers, and 
all O atoms are fully relaxed until the forces on atoms converge to less 
than 0.02 eV/\AA.
A dipole correction \cite{NS92,Ben99} was applied to compensate for asymmetry 
of the potential at the two sides of the slab.
To  sample the Brillouin zone (BZ), Monkhorst-Pack {\bf k}-point meshes
\cite{MoPa76}  were applied, corresponding to $20\times 12 \times 1$
{\bf k}-points in the $1 \times 1$ primitive unit cell, and accordingly 
reduced for larger cells.
To improve the quality of BZ integrations for the fractional occupancies
a Methfessel-Paxton method \cite{MetPa89} with a broadening of 0.2 eV is
applied. In the calculations for slabs the spin-polarization effects were
not included.
The binding energy of oxygen is calculated with respect to one-half of the
energy of a spin-polarized O$_2$ molecule placed in a large rectangular 
unit cell.

Our slab reproduces well \cite{KieNie04} both the electronic properties  
(the work function equals to 4.12 eV) and the geometric structure of the 
upper layers of bare Mo(112), including a 17\% contraction of the interplanar 
distance for the topmost atomic plane, and a relatively small vertical 
relaxation of the second ($-0.9$\%), third (3.0\%), and fourth ($-0.3$\%)
interlayer spacing. 

\section{RESULTS AND DISCUSSION}
\subsection{O adsorption on unreconstructed Mo(112)}

\begin{figure}[b]
\scalebox{0.6}{\includegraphics{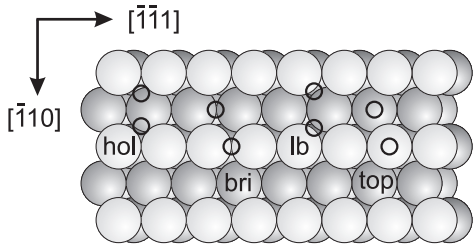}}
\caption{Top view of the upper layers of the ideal furrowed  Mo(112) surface 
and the possible locations of adsorption sites for O atom.
From left to the right two possibilities for quasi threefold hollow
(hol), short bridge (bri), long bridge (lb), and atop (top) sites are 
represented by black circles. \label{f1}
}
\end{figure}

In order to determine the preferred sites for O adsorption on a clean Mo(112) 
we consider first very low coverages of 1/8 ML corresponding to a single
O atom in a $4\times 2$ surface unit cell. We define the coverage $\Theta$ 
as the ratio of the number of adatoms to the number of substrate atoms in a 
surface unit cell.
Following experimental findings \cite{Aruga95,Sas02} we explore several 
possible  adsorption sites for O atoms: long-bridge, atop, quasi-threefold 
hollow, and short-bridge.
The locations of these sites are shown schematically in Fig.~\ref{f1}.
%%------------------------------------------------------------------
%%------------------------------------------------------------------

At this coverage the distances between adatoms are large (10 \AA{} across,
and 11 \AA{} along the rows), and the O--O interaction is negligible.
The strongest binding for this coverage is found when the O atoms are
adsorbed in short bridge sites in a topmost Mo row (Table~\ref{t1}).
Note that the binding (adsorption) energy, which is calculated with respect
to the energy of one-half of the energy of a free O$_2$ molecule, from the
total energy difference of the systems with and without adsorbate, is defined
here as positive. \cite{KieNie02}
A second stable site at low coverages is the quasi-threefold hollow site 
situated between two row atoms and one trough atom with the binding energy 
about 120 meV lower. In the following, for brevity, the former sites are 
referred to as bridge and the latter as hollow sites.
The other sites appear either unstable (long-bridge in the trough, and atop
Mo atom in the rows --- the O atoms shift from there to hollow or to
short-bridge sites, respectively) or show a distinctly weaker binding (atop
is 0.6 eV weaker bound) compared to the strongest bound O in short-bridge
sites.
Interestingly, the in-trough located short-bridge sites are equally unfavored
as on-top sites. Actually, they both end up in a shifted short-bridge site
which could be called a hollow site of second type. This site is coordinated
by one first-layer atom and two second layer Mo atoms.
A preference for the bridge sites on the top Mo row is somewhat
unexpected because oxygen atoms usually tend to occupy a higher coordinated
sites on transition metal surfaces. Therefore, it contradicts the experimental
models which suggest a quasi-threefold hollows as most favored sites for
low-coverage O-adsorption on Mo and W(112). \cite{Fuku93,Aruga95,Bu89}
It also differs from O adsorption on the furrowed Cu(110) surface where both
experiment and theory \cite{Lie98} predict high-coordinated hollow sites as
most favored.
%%%%%%%%%%%%%%%%%%%%%%%%%%%%%%%%%%%%%%%%%%%%%%%%%%%%%%%%%%%%%%%%%%%%%%%%%
\begin{table}
\caption{\label{t1} Binding energies of O atom adsorbed in
short bridge positions at a topmost Mo rows and quasi-threefold hollows
of Mo(112) in different adsorbate structures. The last line entry corresponds 
to the bottom-right $(1\times 2)$ pairing-row structure
displayed in Fig.~\ref{f2}.
}

\begin{ruledtabular}
\begin{tabular}{cccc}
Coverage [ML] & Structure & \multicolumn{2}{c}{Binding energy [eV]} \\
 \cline{3-4}
             &           &   bridge & hollow           \\
\hline
%----------------------------------------------------------
 0.125       & $(4\times 2)$ & 	  4.322 &	4.195  \\ \hline
%----------------------------------------------------------
  0.25       & $(2\times 2)$ & 	  4.305 & 	4.148  \\ \hline
%----------------------------------------------------------
  0.5        & $(2\times 1)$  &	  4.257 &	4.130  \\
             & $c(2\times 2)$ &	  4.234 &	4.145  \\
             & $(1\times 2)$  &	  3.691 &	3.921  \\  \hline
%----------------------------------------------------------
 1           & $(1\times 1)$  &     3.566 &	3.907  \\
             & $(1\times 2)$  &           &	3.906  \\
%----------------------------------------------------------
\end{tabular}
\end{ruledtabular}
\end{table}
%%%%%%%%%%%%%%%%%%%%%%Table~\ref{t1}%%%%%%%%%%%%%%%%%%%%%%%%%%%%%%%

Table~\ref{t1} presents the calculated binding energies.
The O coverage was increased either by placing additional O atoms in the
surface unit cell or by reducing cell size. 
Test calculations performed at the coverage $\Theta=0.5$ for the 
bridge- and hollow-site $p(1\times 2)$-O structures in $1\times 2$ and 
$2\times 2$ unit cells, show that the binding energies agree within 15 meV 
and 13 meV, respectively. Hence, the maximum error bar in the determination 
of the relative energies, with respect to different size of unit cells and
different number of {\bf k}-points applied, is estimated as $\pm 15$ meV.
For a given coverage, we compare the energies of different patterns 
calculated in a unit cell of same size and the error is much below this 
limit. 
The binding energy differences for low-coverage structures amount to at 
least several tens of meV, thus they are well beyond the above error bars 
and different structures can be easily discriminated.

\begin{figure}%[b]
\scalebox{0.44}{\includegraphics{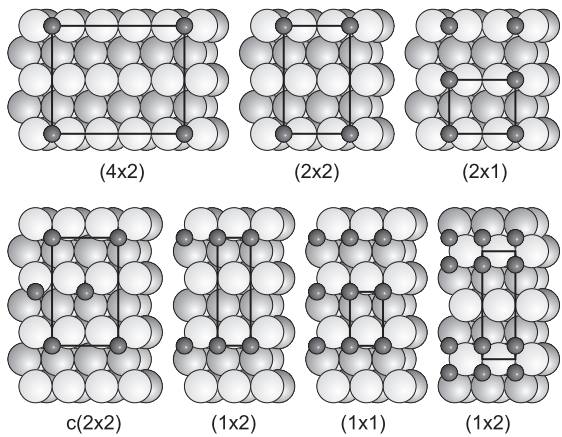}}
\caption{\label{f2} Schematic view of the adatom structures formed at
various O coverages on the Mo(112) surface.
The configurations of the upper row correspond to oxygen adsorbed in bridge
sites while those of the lower row to O adatoms in hollow sites.
For all the displayed O configurations both types of adsorption sites
are considered (Table~\ref{t1}) except of the bottom-right pairing-row
structure which appears only for the hollow sites.
}
\end{figure}

The results of Table~\ref{t1} seem to agree with the scenario of an initial 
stage of O adsorption based on the low energy electron diffraction (LEED) 
experiment. \cite{Fuku93}
According to that, beginning from lowest coverages, the oxygen atoms tend 
to form atomic rows
along the $[\overline{1}\overline{1}1]$ direction with twice the spacing
between atoms in the close-packed Mo rows. The distance and phase of adjacent
O rows are initially random. At increased coverage O-rows tend to occupy
in-phase every second row. This results in formation of a $(2\times2)$ pattern
at 0.25 ML (Fig.~\ref{f2}). At 0.5 ML coverage, the sites along each Mo row
are occupied and the $p(2\times1)$-O structure is developed. This pattern
is more stable than the $c(2\times2)$ which is characterized by antiphase
O-rows or the $p(1\times2)$ structure with O atoms in each site along 
every second row. 
It is also clear that for coverages $\Theta \leq 0.5$ ML the bridge
sites remain most favored.

%%------------------------------------------------------------------
%%------------------------------------------------------------------

With increased coverage ($\Theta=0.5$), when the O atoms begin to occupy
neighboring sites along the rows [$p(1\times2)$ structure], quasi-threefold
hollow sites become more preferred than bridge sites (Table~\ref{t1}).
Note, however, that for this coverage the $p(2\times1)$ and $c(2\times2)$
structures with a more homogeneous distribution of O atoms still remain
energetically most favored. These results suggest that O--O repulsion is
stronger for bridge than for hollow adsorption. In the latter positions
the O adatoms are immersed in the electron charge distribution of the Mo
substrate which screens the dipole--dipole repulsion. For a higher coverage
our results predict a structural phase transition, from a $p(1\times1)$
phase with adatoms preferentially occupying the bridge sites, to a
$p(1\times1)$ structure where adatoms occupy hollow sites.
The energetic differences for O adsorbed in hollow sites at different
coverages and structures are smaller than for O in bridge-top sites.

For both types of energetically favorable sites a decrease in the binding
energy with increasing coverage is clearly seen (Table~\ref{t1}).
While for hollow sites the changes in the binding energy do not exceed
190 meV, for the whole range of coverages considered they are four times
larger for bridge sites.
In the latter case, they increase to 0.7 eV for $\Theta \geq 0.5$ in the
cases when the nearest neighbor sites parallel to the troughs are occupied.
For bridge sites the stability of their structures enhances with aligning
adatoms normal to the furrows whereas for hollow sites the opposite can be
observed.

Our results show (Table~\ref{t1}) that the $(1\times2)$-O pairing-row
reconstruction, illustrated in the bottom-right panel in Fig.~\ref{f2},
where O atoms occupy hollow sites on both sides of the Mo row is equally
favored as the $(1\times1)$ structure with O atoms homogeneously distributed
in hollow sites on one side of the Mo rows. In LEED experiments one originally
observes $(1\times1)$ patterns, and a $(1\times2)$ pattern is formed only
after annealing to ca.\ 500--600 K. \cite{Fuku93,Aruga95}
The calculated barrier (0.63 eV) for O diffusion from a hollow site on one 
side of the trough to the one located on the other side (Fig.~\ref{f1}) 
confirms that this configuration can be reached only during activated 
adsorption process. The stability of this structure means that atoms of 
a Mo row screen very effectively the repulsive interaction between O atoms
of the pairing rows.

%%%%%%%%%%%%%%%%%%%%%%%%%Table~\ref{t2}%%%%%%%%%%%%%%%%%%%%%%%%%%%%%%%%%%%
\begin{table}
\caption{\label{t2}
The geometry of oxygen adsorption on Mo(112).
$\Delta d_{12}$ is the change in the vertical relaxations of the topmost
interlayer distance, $h_{1(2)}$ is the height of O atom with respect to 
the topmost (second) Mo layer, $l$ is the O-Mo bond length (distance).
For a clean Mo(112) $\Delta d_{12}=-17.0$\%.
}

\begin{ruledtabular}
\begin{tabular}{cccccc}
Coverage [ML] & Structure  & $\Delta d_{12}$ [\%] & $h_1$ [\AA] & $h_2$ [\AA]
& $l$ [\AA] \\ \hline
\multicolumn{6}{c}{short-bridge}
\\ \hline
0.125 & $(4\times 2)$ &   $-$14.4    & 1.45  &  2.56  &  1.96  \\
0.25  & $(2\times 2)$ &   $-$12.0    & 1.46  &  2.59  &  1.94  \\
0.5   & $(2\times 1)$ &   $-$7.1     & 1.37  &  2.56  &  1.95  \\
      & $(1\times 2)$ &   $-$8.7     & 1.51  &  2.68  &  1.95  \\
1     & $(1\times 1)$ &   $-$2.0     & 1.37  &  2.63  &  1.92  \\ \hline
\multicolumn{6}{c}{quasi-threefold hollow}
\\ \hline
0.125 & $(4\times 2)$ &   $-$17.1    & 0.87  &  1.94  &  2.03  \\
0.25  & $(2\times 2)$ &   $-$17.9    & 0.86  &  1.92  &  2.03  \\
0.5   & $(2\times 1)$ &   $-$17.4    & 0.87  &  1.93  &  2.05  \\
      & $(1\times 2)$ &   $-$18.1    & 0.79  &  1.85  &  2.07  \\
1     & $(1\times 1)$ &   $-$20.8    & 0.81  &  1.83  &  2.06  \\
\end{tabular}
\end{ruledtabular}
\end{table}
%%%%%%%%%%%%%%%%%%%%%%%%%%%%%%%%%%%%%%%%%%%%%%%%%%%%%%%%%%%%%%%%%%%%%%%%%%

%%--------------------------------------------------------------------
\begin{figure}[t]
\scalebox{0.5}{\includegraphics{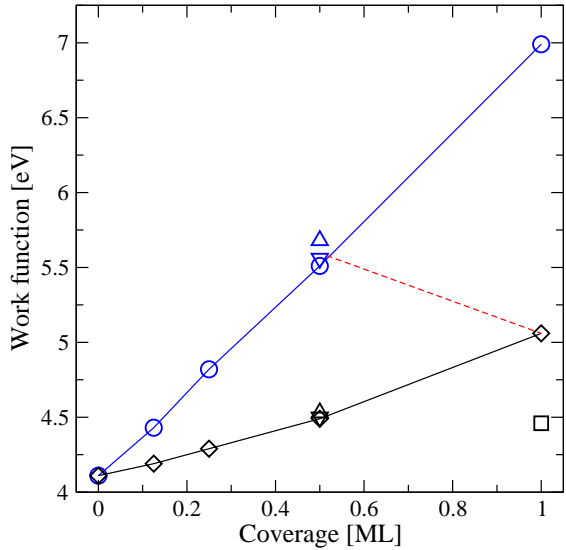}}
\caption{Work function changes induced by O atoms adsorbed on the Mo(112)
surface in short bridge positions (circles) of the first layer and
threefold hollows (diamonds).
The work function for a pairing-row $p(1\times 2)$-O structure is marked
by a square. Triangles down and up (at 0.5 ML) label the entries for the
$c(2\times 2)$ and $(1\times 2)$ structures, respectively. \label{f3}
}
\end{figure}
%%--------------------------------------------------------------------

The changes in the average geometry of Mo(112) layers with O coverage
are presented in Table~\ref{t2}. It is seen that O adatoms very differently
influence the geometry of Mo(112).
The O atoms adsorbed in bridge sites have a stabilizing effect on the Mo(112) 
 surface which is manifested in a diminished contraction (derelaxation) of
the interlayer distance between the topmost Mo layers with increasing
O coverage. In contrast, oxygen adsorbed in the quasi-threefold hollow
sites enhances relaxation of the topmost layers. For higher coverages the
vertical shifts of the atoms of the particular planes relative to the
center-of-mass position (surface rumpling) are much smaller or negligible.
Interestingly, the pairing-row $p(1\times 2)$-O structure involves some
parallel shifts of top Mo rows. Every second Mo row decorated by O atoms
from both sides, shifts in the $[\overline{1}\overline{1}1]$ direction
by about 0.3 \AA{}, whereas the other, ``bare'' Mo-row, not coordinated
directly by oxygen, shifts in an opposite direction by approximately 0.1 \AA.
Furthermore, the bare Mo-rows are shifted down (by $\sim 0.09$ \AA), while
the other with coordinating O-rows, shift up by the same amount with respect 
to the center-of-mass position.
A single O atom adsorbed in a bridge site in a $4\times 2$ cell induces
substantial vertical shifts of the two Mo atoms closest to oxygen (0.1 \AA{}
up). The remaining atoms of the same Mo row do not change their positions.
For an O atom adsorbed in a quasi-threefold hollow, one of the two 
coordinating Mo atoms of the upper row  moves down (0.1 \AA) while the 
other shifts 0.1 \AA\ upwards, and the O atom shifts laterally (by 0.08 \AA) 
in the direction of the latter. All these adjustments enhance the 
coordination of the oxygen atom.
No change in the vertical alignment of the Mo rows is observed.
The decrease in the binding energy with coverage is, in general, 
accompanied by a small, though not regular, decrease in the vertical 
distance of the O adatom to the topmost Mo plane.

The distance $h_2$ to the second Mo layer increases with coverage for bridge sites and decreases for hollow sites. For the bond length $l$ an opposite 
trend is observed. Variations of $l$ correlate with a reduced contraction 
of the topmost Mo-layers spacing. 

The calculated work function variation due to adsorption of electronegative 
O exhibits a strong increase with coverage which is consistent with a
general picture of oxygen chemisorption on transition metals.
Note that this increase is almost three times larger for O atoms adsorbed
in short-bridge positions than for those in the hollow sites.
This is presumably due to a much stronger polarization of protruding
O adatoms in bridge than hollow sites.
The work function variation seems to be consistent with experiment which
reports a rapid increase of work function followed by a slower decrease 
for higher oxygen exposures. \cite{Schro04}
Our binding energy data (Table~\ref{t1}) show that for $0.5 <$ coverages
$< 1$ ML a site exchange transition from bridge to hollow sites occurs,
therefore a work function decrease marked schematically in Fig.~\ref{f3}
by a dashed line should accompany this transition.

%%-------------------------------------------------------------------------
\begin{figure}[t]
\scalebox{0.5}{\includegraphics{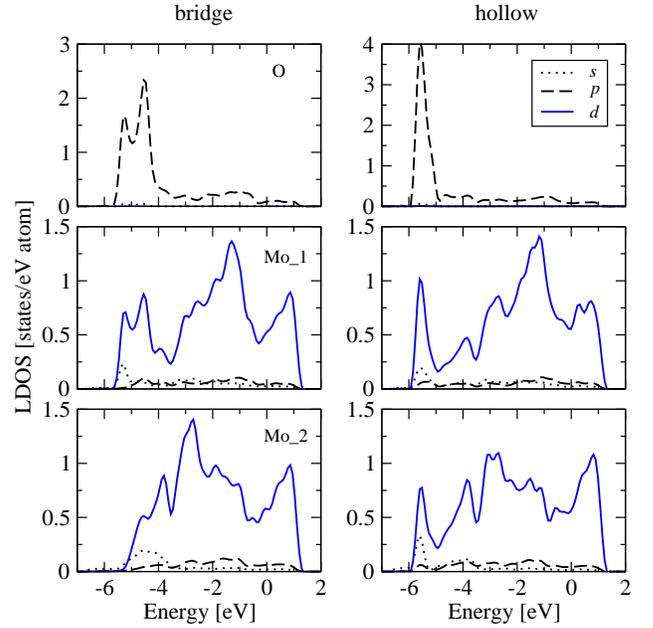}}
\caption{\label{f4}
Atom resolved density of states for O atom and the nearest Mo atoms of the
topmost two layers of Mo(112) at 0.25 ML of oxygen coverage. The left column
displays the states for O adsorbed in bridge site while the right one in
hollow site. The energy is measured with respect to the Fermi level.
}
\end{figure}
\begin{figure}[t]
\scalebox{0.5}{\includegraphics{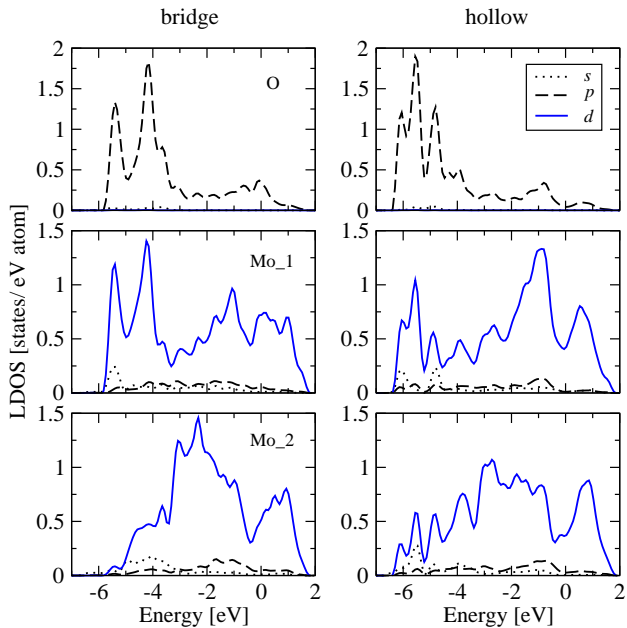}}
\caption{Same as in Fig.~\ref{f4} for 1 ML coverage.
\label{f5}
}
\end{figure}

%%-------------------------------------------------------------------------
Figures ~\ref{f4} and \ref{f5} display the local partial density
of electronic states (LDOS) at the O atom and the neighboring atoms of
the two topmost layers, for a low and a full monolayer coverage with O
atoms in bridge and hollow sites.
The main contribution to the bonding comes from the O $2p$ orbitals. 
For 1/4 ML coverage, with O in bridge sites this contribution consists
of two partly overlapping peaks, with the higher of them located at $-4.5$ eV.
For O in hollow site the region of $2p$ states bonding is narrower
($\simeq 1$ eV wide) with only one peak centered at $-5.5$ eV. As is seen
from Fig.~\ref{f4}, the O $2p$ orbitals hybridize strongly with the Mo $4d$
states. For oxygen adsorbed in the bridge site the bonding is limited only to 
the Mo atom of the topmost row while for O adsorbed in hollow sites both the 
Mo atom of the upper row and the one of the trough contribute to the bonding. 
The density of states of the Mo atom of the trough resembles very much that 
of Mo bulk \cite{Moruzzi} which is dominated by a conduction band formed of localized $4d$ orbitals.

%%-------------------------------------------------------------------------
%%-------------------------------------------------------------------------

For a 1 ML coverage (Fig.~\ref{f5}), the regions of bonding widens both for
bridge and hollow sites adsorption. It extends from $-5.8$ to $-3.5$ eV with
two peaks at about $-5.5$ and $-4.2$ eV, for the bridge site adsorption.
Again, the O $2p$ orbitals hybridization with the Mo $4d$ orbitals contributes mostly to the bonding.
For O adsorption in hollow sites, the single peak due to O $2p$ orbitals
is split into three distinct components and the range of bonding has
a similar width to that for adsorption in bridge sites. The hybridization
of the $5s$ states of the upper Mo row with O $2p$ orbitals is also increased.

\subsection{Missing-row reconstruction of Mo(112)}

\begin{figure}
\scalebox{0.5}{\includegraphics{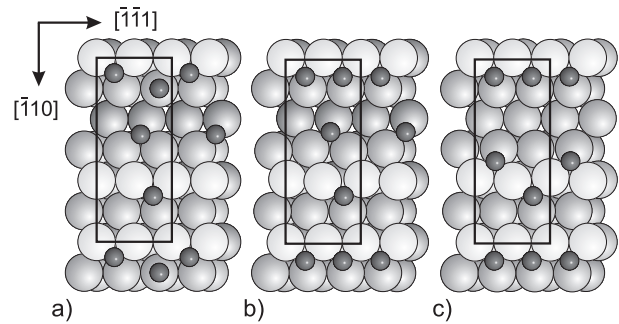}}
\caption{A $p(2\times 3)$ unit cell of the missing row reconstruction of
the Mo(112) surface induced by O adsorption. (a) experimental model, (b)
structure optimized positions of model (a), an example of the calculated 
more stable structure (c). Oxygen atoms are represented by small dark spheres.
\label{f6}
}
\end{figure}

It has been recently reported that O adsorption induces a $p(2\times 3)$
surface reconstruction of the missing row type. \cite{Schro02,Schro04}
A structure model of reconstruction was proposed [Fig.~\ref{f6}(a)], 
resulting from a combined scanning tunneling microscopy and LEED study. 
In order to check stability of this structure, which assumes oxygen placed 
both in hollow and on-top Mo-atom positions, we have used it as an input 
model for the total energy calculations. 
%%-------------------------------------------------------------------------
%%-------------------------------------------------------------------------
The structure optimization leads to the O-pattern [Fig.~\ref{f6}(b)] 
which differs from the one suggested by experiment. \cite{Schro02,Schro04}
In the optimized structure, the oxygen atom placed originally atop the 
Mo atom of the second Mo layer is shifted to the quasi-threefold hollow, 
thus aligning the O atoms with a Mo row.
This is not too surprising in view of our findings, discussed in Sec.~IIIA,
that on-top sites belong to the least favored ones for O adsorption at Mo(112) and are 0.6 eV less stable than hollow sites.
It means that an O atom sitting in a hollow site provides a more effective
coordination to the Mo atoms of second layer than atop configuration.

The calculated $p(2\times 3)$ structure belongs to the most stable 
ones among several considered configurations of O atoms in the unit cell, 
but even limited search for the total energy minimum, within a class of 
different O assignments, allows to find other, more stable structures, 
involving oxygen atoms of only hollow sites. An example of such a structure 
(prefered by 446 meV per unit cell) is given in Fig.~\ref{f6}(c). 
Thus, though the optimized structure of Fig.~\ref{f6}(b) corresponds best 
to the one suggested by the experimental studies \cite{Schro02,Schro04} 
it does not provide a global energy minimum and should be considered as 
one of the most likely but metastable configurations.

\section{Conclusions}

The systematic total-energy calculations of atomic oxygen adsorption
on the relaxed Mo(112) surface are presented and discussed.
In contrast to experimental assignments our calculations identify the
short-bridge sites of the uppermost Mo row as the most stable for O
adsorption at lower coverages.
They are by about 120--150 meV per atom more stable than the quasi-threefold
hollows. The binding energy of O atoms adsorbed in the stable bridge and
quasi-threefold hollow sites decreases with increasing coverage.
Our results show that O atoms very differently modify the spacing between 
the topmost two Mo(112) layers. Thus, measurements of lattice relaxation 
can provide a method for identification of the adsorbed oxygen sites.
At one monolayer coverage a site exchange transition from bridge to
quasi-threefold hollow sites occurs.
The calculated work function variation with O adatom coverage is found to
be consistent with recent measurements.
The stability of the $p(1\times 2)$ pairing-row reconstructed O pattern
predicted by LEED experiment is confirmed.
The $p(2\times 3)$ missing-row reconstruction is investigated. 
The stability of the configuration suggested by experiment requires 
slightly modified positions of O atoms.

\begin{acknowledgments}
The work of A.K. was supported  by the Polish Committee for Scientific
Research (KBN) grant No.\ 1 P03B 047 26.
We acknowledge support from the Academy of Finland through the Center of
Excellence Program (2000-2005) and the generous allocation of computer
resources of the Centre for Scientific Computing (CSC), Espoo, Finland.
\end{acknowledgments}

%\newpage
%%%%%%%%%%%%%%%%%%%%%%%%%%%%%%%%%%%%%%%%%%%%%%%%%%%%%%%%%%%%%%%%%%%%%


\begin{thebibliography}{99}

\bibitem{Schro02}
T. Schroeder, J. B. Giorgi, A. Hammoudeh, N. Magg, M. B\"aumer, and
H.-J. Freund, Phys. Rev. B \textbf{65}, 115411 (2002).

\bibitem{Schro02a}
T. Schroeder, J. B. Giorgi, M. B\"aumer, and H.-J. Freund,
Phys. Rev. B \textbf{66}, 165422 (2002).

\bibitem{Chen04}
M.S. Chen, A.K. Santra, and D.W. Goodman,
Phys. Rev. B \textbf{69}, 155404 (2004).

\bibitem{KieNie02}
A. Kiejna and R.M. Nieminen, Phys. Rev. B \textbf{66}, 085407 (2002); 

\bibitem{KieNie04}
A. Kiejna and R.M. Nieminen, Phys. Rev. B \textbf{69}, 235424 (2004).

\bibitem{KoPf_2}
D. Kolthoff and H. Pfn\"ur, Surf. Sci. \textbf{459}, 265 (2000).

\bibitem{Godz04}
G. Godzik, T. Block, and H. Pfn\"ur, Phys. Rev. B \textbf{69}, 235414 (2004).

\bibitem{Godz01}
G. Godzik, H. Pfn\"ur, and I. Lyuksyutov, Europhys. Lett. \textbf{56}, 67 (2001).

\bibitem{Fuku93}
K. Fukui, T. Aruga, and Y. Iwasawa, Surf. Sci. \textbf{281}, 241 (1993).

\bibitem{Aruga95}
T. Aruga, K. Tateno, K. Fukui, and Y. Iwasawa,
Surf. Sci. \textbf{324}, 17 (1995).

\bibitem{Sas02}
T. Sasaki, Y. Goto, R. Tero, K. Fukui, and Y. Iwasawa,
Surf. Sci. \textbf{502-503}, 136 (2002).

\bibitem{Ertl67}
G. Ertl, Surf. Sci. \textbf{6}, 208 (1967).

\bibitem{Lie98}
S.Y. Liem, G. Kresse, and J.H.R. Clarke, Surf. Sci. \textbf{415}, 194 (1998).

\bibitem{McA00}
T. McAvoy, J. Zhang, C. Waldfried, D.N. McIlroy, P.A. Dowben, O. Zeybek,
T. Bertrams, S.D. Barrett,
 Eur. Phys. J. B \textbf{14}, 747 (2000).

\bibitem{Schro04}
T. Schroeder, J. Zegenhagen, N. Magg, B. Immaraporn, and H.-J. Freund,
Surf. Sci. \textbf{552}, 85 (2004).

\bibitem{HK64}
P. Hohenberg and W. Kohn, Phys. Rev. \textbf{136}, B864 (1964).

\bibitem{KS65}
W. Kohn and L. J. Sham, Phys. Rev. \textbf{140}, 1133A (1965).

\bibitem{Per92}
J. P.\ Perdew, J. A. Chevary, S. H.\ Vosko, K. A.\ Jackson, M. R.\ Pederson,
D. J.\ Singh, and C.\ Fiolhais,
Phys. Rev. B \textbf{46}, 6671 (1992); \textbf{48}, 4978 (1993).

\bibitem{Blo94}
P. Bl\"ochl, Phys. Rev. B \textbf{50}, 17953 (1994).

\bibitem{paw}
G. Kresse and D. Joubert, Phys. Rev. B \textbf{59}, 1758 (1999).

\bibitem{vasp}
G. Kresse and J. Hafner, Phys. Rev. B \textbf{47}, R558 (1993);
\textbf{49}, 14 251 (1994); G. Kresse and J. Furthm\"uller,
Phys. Rev. B \textbf{54}, 11 169 (1996); Comput. Mater. Sci. \textbf{6},
15 (1996).

\bibitem{NS92}
J.\ Neugebauer and M.\ Scheffler, Phys. Rev. B \textbf{46}, 16067 (1992).

\bibitem{Ben99}
L. Bengtsson, Phys. Rev. B \textbf{59}, 12301 (1999).

\bibitem{MoPa76}
H. J. Monkhorst and J. D. Pack, Phys. Rev. B \textbf{13}, 5188 (1976).

\bibitem{MetPa89}
M.\ Methfessel and A. T.\ Paxton, Phys. Rev. B \textbf{40}, 3616 (1989).

\bibitem{Bu89}
H. Bu, O. Grizzi, M. Shi, and J. W. Rabalais, 
Phys. Rev. B \textbf{40}, 10147 (1989).

\bibitem{Moruzzi}
V.L. Moruzzi, J.F. Janak, and A.R. Williams, 
\textit{Calculated Electronic Properties of Metals}
(Pergamon Press, New York, 1978).

\end{thebibliography}
\end{document}